# Introduction to new demographic model for humans

Byung Mook Weon

*LG.Philips Displays, 184, Gongdan1-dong, Gumi-city, GyungBuk, 730-702, Korea*

*Author for correspondence (bmw@lgphilips-displays.com)*

The Gompertz model since 1825 has significantly contributed to interpretation of ageing in biological and social sciences. However, in modern research findings, it is clear that the Gompertz model is not successful to describe the whole demographic trajectories. In this letter, a new demographic model is introduced especially to describe human demographic trajectories, for example, for Sweden (2002). The new model is derived from the Weibull model with an age-dependent shape parameter, which seems to indicate the dynamical aspects of biological systems for longevity. We will discuss the origin of the age-dependent shape parameter. Finally, the new model presented here has significant potential to change our understanding of definition of maximum longevity.

Fundamental studies of the ageing process have lately attracted the interest of researchers in a variety of disciplines, linking ideas and theories from such diverse fields as biochemistry to mathematics (Weitz & Fraser 2001). In general, the way to characterize ageing is to plot the increase in mortality rate with chronological age. The fundamental law of population dynamics is the Gompertz law (Gompertz 1825), in which the human mortality rate increases roughly exponentially with increasing age at senescence. The Gompertz model is most commonly employed to compare mortality rates between different populations (Penna & Stauffer 1996). However, no mathematical model so far, including the Gompertz model, has been suggested that can perfectly approximate the dynamical development of the mortality rate over the total life span (Kowald 1999). Particularly in modern research findings, it seems to be obvious that the mortality rate does not increase according to the Gompertz model at the highest ages (Vaupel 1997; Robine & Vaupel 2002), and this deviation from the Gompertz model is a great puzzle to demographers, biologists and gerontologists. Many of the traditional mathematical models (for instance, the Gompertz, Weibull, Heligman & Pollard, Kannisto, Quadratic and Logistic models) for the mortality rate provide poor fits to empirical population data at the highest ages (Thatcher *et al.* 1998; Yi & Vaupel 2003).

In order to describe human demographic trajectories, a new demographic model has been found by Weon for the first time to our knowledge. We would like to call this new model the “Weon model” hereafter, as the Gompertz model was formulated by Benjamin Gompertz and thus it bears his name. In this letter, we wish to introduce the main findings of the Weon model into the scientific societies for biology.

For analysis, we used the life tables by year of death (period, for all sexes, 1x1) for 2002 for Sweden, which is one of the longest-lived countries, from the Human Mortality Database (www.mortality.org). The survival probability is expressed as a fraction ($l_x/l_o$) of the number of survivors ($l_x$) out of 100,000 persons ($l_o$) in the original life tables.

For statistical definitions, let $f(t)$ be the *probability density function* (*pdf*) describing the distribution of life spans in a population. The *cumulative density function* (*cdf*), $F(t)$, gives the probability that an individual dies before surpassing age $t$ (especially age $t$ is a *continuous random variable*). The *survival function*, $S(t)$, gives the complementary probability ($S(t) = 1 - F(t)$) that an individual is still alive at age $t$. The *mortality function*, $\mu(t)$, is defined as the ratio of the density and survival functions ($\mu(t) = f(t)/S(t)$). Thus, the mortality function gives the probability density at age $t$ conditional on survival to that age.

The Weon model is derived from the Weibull model (Weibull 1951) with an assumption that the shape parameter is a function of age. In the Weibull model, the shape parameter is constant with age (Nelson 1990). The age-dependent shape parameter enables us to model the demographic (survival and mortality) functions, which are expressed as follows,

$$S(t) = \exp(-(t/\alpha)^{\beta(t)})$$

$$\mu(t) = (t/\alpha)^{\beta(t)} \times [\frac{\beta(t)}{t} + \ln(t/\alpha) \times \frac{d\beta(t)}{dt}]$$

where $S(t)$ is the survival function, indicating the probability that an individual is still alive at age $t$ and $\mu(t)$ is the mortality function, indicating the probability density at age conditional on survival to that age, in which $\alpha$ denotes the characteristic life ($t = \alpha$ when $S = \exp(-1) \approx 36.79\%$) and $\beta(t)$ denotes the shape parameter as a function of age. The original idea was obtained as follows: typical human survival curves show i) a rapid decrease in survival in the first few years of life and then ii) a relatively steady decrease and then an abrupt decrease near death. Interestingly, the former behaviour resembles the Weibull survival function with $\beta < 1$ and the latter behaviour seems to follow the case of $\beta >> 1$. With this in mind, it could be assumed that shape parameter is a function of age.

We could evaluate the age dependence of the shape parameter to determine an adequate mathematical expression of the shape parameter, after determination of the characteristic life ($\alpha$) graphically in the survival curve (figure 1 (a)). Conveniently, the value of $\alpha$ is always found at the age for the survival to be '$\exp(-1)$'; this is known as the characteristic life. This feature gives the advantage of looking for the value of $\alpha$ simply by graphical analysis of the survival curve. For Sweden (2002), the value of $\alpha$ was observed to be 86.10 years.

In turn, with the value of $\alpha$, we could plot the shape parameter as a function of age by the mathematical equivalence of '$\beta(t) = \ln(-\ln S(t)) / \ln(t/\alpha)$' (figure 1 (b)). If $\beta(t)$ is not constant with age, this proves that '$\beta(t)$ is a function of age'. Evidently, we see the age dependence of the shape parameter for humans through the trajectory of shape parameter (figure 1 (b)). In empirical practice, we could use a polynomial expression for modelling the shape parameter as a function of age as follows:

$\beta(t) = \beta_0 + \beta_1 t + \beta_2 t^2 + \ldots$, where the associated coefficients could be determined by a regression analysis in the plot of $\beta(t)$. And thus, the derivative is obtained as follows: $d\beta(t)/dt = \beta_1 + 2\beta_2 t + \ldots$, which indicates again that the shape parameter is a function of age. On the other hand, the value of $\beta(t)$ mathematically approaches infinity as the age $t$ approaches the value of $\alpha$ or the denominator '$\ln(t/\alpha)$' approaches zero. This feature leaves the 'trace of $\alpha$' in the plot of $\beta(t)$, thus we can observe variations of $\beta(t)$ and $\alpha$ at once (figure 1 (b)). If $\beta(t)$ (except for the mathematical singularity or the trace of $\alpha$) can be expressed by an adequate mathematical function, the survival and mortality functions can be calculated by the mathematically expressed $\beta(t)$.

On the other hand, it seems that the density function, $f(t) = S(t) \times \mu(t)$, indicates the complementarity between survival and mortality functions. That is, the mortality rate tends to increase with decreasing the survival rate. For longevity, an individual is likely to try to reduce the mortality rate but strive to improve the survival rate. Interestingly, there exists the maximum value (or the peak) of the density functon approximately at the characteristic life (figure 1 (c)).

In practice, a linear expression for $\beta(t)$ is roughly appropriate for ages before $\alpha$ and a quadratic expression is appropriate for ages after $\alpha$ (figure 1 (b)). These mathematical expressions of the shape parameter can solve the conventional question as follows: the mortality rate does not increase according to the Gompertz model at the highest ages (80+), and this deviation from the Gompertz model is a great puzzle (Vaupel 1997). It is still not certain whether the mortality trajectories level or decrease at the highest ages (Robine & Vaupel 2002). Some researchers have suggested the mortality curves tending but never reaching a plateau or a ceiling of mortality (Thatcher *et al.* 1998; Thatcher 1999; Lyncher & Brown 2001), whereas others have suggested that the mortality curves could decrease after having reached a maximum (Vaupel *et al.* 1998; Robine & Vaupel 2002). Specifically, the mortality trajectories for higher ages (110+) are essential to understand the maximum longevity for humans. If the quadratic expression of the age-dependent shape parameter for ages 80-109 is still valid for ages 110+, the Weon model enables us to predict that the mortality rate will decline after a plateau around ages 110-115 and the maximum longevity emerges around ages 120-130 for the modern demographic data (figure 1 (d)). Furthermore, through an approximate relationship of '$\ln\mu \propto \beta(t)$' at middle age (~30-80) by a linear expression for $\beta(t)$, the Weon model can approximate the Gompertz model when '$\beta(t) \propto t$' (figure 1 (b) and a small figure in 1 (d)). Particularly, the mortality rate deviates from the Gompertz model when $\beta(t)$ has a non-linear behaviour at the highest ages (80+). The Weon model can therefore generalize the Gompertz model as well as the Weibull model: That is, the Gompertz model is a special case of a linear expression for $\beta(t)$ and the Weibull model is a special case of a constant shape parameter.

By the way, what is the origin of the age-dependent shape parameter? It may indicate the dynamical aspects of biological systems for longevity. In principle for the highest value of the survival function or for longevity at all times, the shape parameter should be variable dynamically according to the characteristic life: "For longevity, $\beta(t)$ tends to increase at $t < \alpha$ but it tends to decrease at $t > \alpha$." This is attributable to the nature of biological systems to strive to survive healthier and longer robustly against intrinsic defects and circumstances. The quadratic coefficient in a parabola for $\beta(t)$ could quantitatively measure the decrease (or bending down) of $\beta(t)$ at $t > \alpha$ (figure 1 (b)). Interestingly, because of the reason of longevity (something to cause decreasing $\beta(t)$ at $t > \alpha$), the limit of longevity inevitably seems to emerge (figure 1 (d)). We call this phenomenon the "complementarity of longevity". That is, the longevity tendency bends down the shape parameter after the characteristic life, which simultaneously causes the emergence of the limit of longevity. It is the age dependence of the shape parameter that seems to be governed by the complementarity of longevity.

Finally, the Weon model presented here has significant potential to change our understanding of definition of the maximum longevity. In general, the term of "*longevity*" means the "*duration of life*". In a sense, the "*maximum longevity*" can be used to mean the "*maximum duration of life*" of a given population. However, what we know is the "*maximum age at death*", which means the oldest age at death observed in a given population during a given time period (Vallin & Meslé 2001). Perhaps the most common notion of a limit in the study of human longevity is the *limited-life-span hypothesis*, which states that there exists some age ($\omega$) beyond which there can be no survivors. This hypothesis can be expressed by any one of the following three formulas (Wilmoth 1997): "$f(t) = 0$, $S(t) = 0$ or $\lim_{t \to \omega} \mu(t) = \infty$ ($t \geq \omega$)." However, according to the Weon model, the survival function may be not zero, although it has extremely low values at the highest ages, while the mortality function can be zero at the maximum longevity (figure 1 (d)). The Weon model suggests that the maximum longevity can be defined as follows: "at $t = \omega$, $f(t) = 0$ and $\mu(t) = 0$, instead of $S(t) = 0$." Fundamentally, the decrease rate of the survival function with age ($-dS(t)/dt$; this term means the density function (*pdf*), $f(t)$, and the minus indicates the decrease) should be zero at the maximum longevity. Therefore, the maximum longevity can be simply defined as "$-dS(t)/dt = 0$ or $\mu(t) = 0$". In practice, we can identify the maximum longevity at the moment that the survival trajectory levels off, or the mortality trajectory becomes zero (figure 1 (d)).

**Acknowledgements**

The author thanks to the Human Mortality Database (Dr. John R. Wilmoth, as a director, in The University of California, Berkeley and Dr. Vladimir Shkolnikov, as a co-director, in The Max Planck Institute for Demographic Research) for allowing anyone to access the demographic data for research.

**Figure legend**

**Figure 1. Demographic trajectories for Sweden (2002) using the Weon model.** (a) Survival function: survival data from the life tables (dot). In this plot, the characteristic life could be graphically measured at the survival of 'exp(-1)'. (b) Shape parameter: quasi-data calculated by survival data and estimated characteristic life (dot), and mathematical model (a parabola) of shape parameter after characteristic life through a regression analysis (line). (c) Density function (*pdf*): data multiplied by survival and mortality data (dot), and model fit (line) by modelling shape parameter. Empirically, the maximum density (or the peak) exists at the characteristic life. (d) Mortality rate: mortality data (dot), and model fit (line) by modelling shape parameter. The mortality trajectory is predicted to decline after a plateau and reach zero at a maximum longevity, which is directly related to the density function to be zero at that point. Interestingly, the Gompertz model (showing a simple exponential increase of mortality rate with age) is useful merely at the middle age (~30-80 in a small figure in d), which is associated with a linear-like trend of shape parameter in figure (b). The Weon model by mathematical modelling of the age-dependent shape parameter is more flexible than the Gompertz model to describe dynamical demographic trajectories over the total life span.

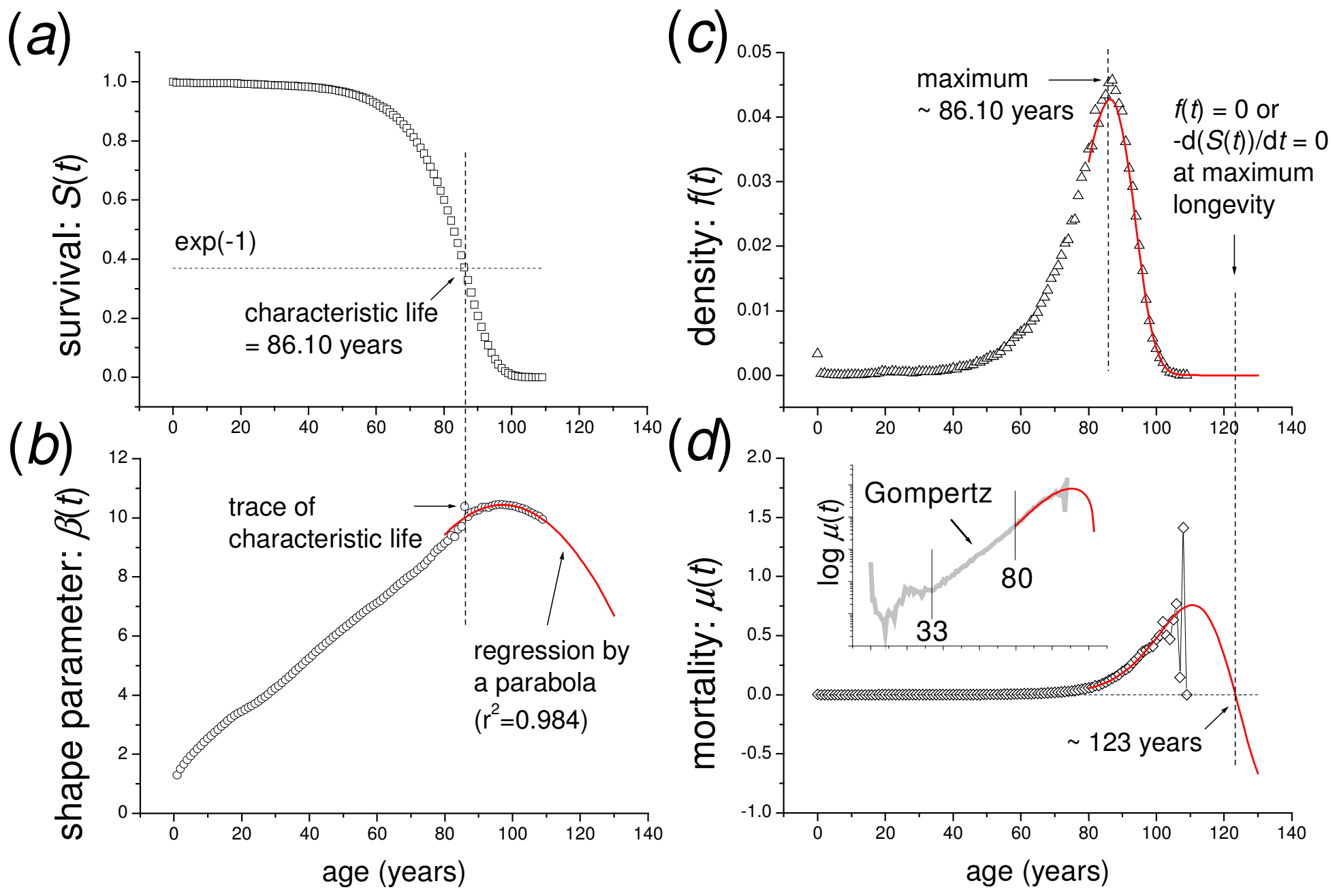

**Figure 1.**